\documentclass[conference, 9pt]{IEEEtran}
\ifdefined\blind
\newcommand\githublink{\url{https://github.com/anonymous/repository}}
\else
\newcommand\githublink{\url{https://github.com/joennlae/halutmatmul}}
\fi

\newcommand\isvlsi[1]{\textcolor{black}{#1}}
\IEEEoverridecommandlockouts
\usepackage{cite}
\usepackage{amsmath,amssymb,amsfonts}
\usepackage{algorithmic}
\usepackage{graphicx}
\usepackage{textcomp}
\usepackage{xcolor}
\usepackage{fancyhdr}
\usepackage{glossaries}
\usepackage{lipsum}
\usepackage{amsmath} 
\usepackage{commath}
\usepackage{mathtools}
\usepackage{textgreek}
\usepackage[ruled, linesnumbered]{algorithm2e}

\usepackage{array}
\usepackage{arydshln}
\usepackage{tabularx}
\usepackage{hhline}
\usepackage{multicol}
\usepackage{booktabs}
\usepackage{threeparttable}

\definecolor{color1}{RGB}{142,202,230}
\definecolor{color2}{RGB}{33,158,188}
\definecolor{color3}{RGB}{2,48,71}
\definecolor{color4}{RGB}{255,183,3}
\definecolor{color5}{RGB}{251,133,0}
\usepackage{tikz}       
\usetikzlibrary{
  arrows,
  positioning,
  matrix,
  calc,
  decorations.pathreplacing,
  decorations.pathmorphing,
  decorations.markings,
  decorations.text,
  shapes,
  backgrounds,
  shadows,
  trees,
  fit,
  snakes,
  patterns,
  mindmap,
  intersections,
  calendar,
  plotmarks,
  spy}

\usepackage{hyperref}
\usepackage[capitalize]{cleveref}

\DeclareMathOperator*{\argmax}{arg\,max}
\DeclareMathOperator*{\argmin}{arg\,min}

\newcommand{\graphnode}[1]{
    \begin{tikzpicture}
        \node[rounded corners=2pt, inner sep=-1pt, fill=#1, minimum size=8pt] at (0,0){};
    \end{tikzpicture}
}

\newcommand{\graphnumbercolor}[2]{
    \begin{tikzpicture}[baseline=(number.base)]
        \node[fill = #1, circle, inner sep=2pt, text=white, outer sep=0pt] (number) at (0, 0, 0) {\footnotesize{#2}};
    \end{tikzpicture}
}

\def\BibTeX{{\rm B\kern-.05em{\sc i\kern-.025em b}\kern-.08em
    T\kern-.1667em\lower.7ex\hbox{E}\kern-.125emX}}

\newacronym[longplural={Scratchpad Memories}]{SPM}{SPM}{Scratchpad Memory}
\newacronym{ACE}{ACE}{AXI Coherent Extensions}
\newacronym{AMM}{AMM}{Approximate Matrix Multiplication}
\newacronym{AMBA}{AMBA}{Advanced Microcontroller Bus Architecture}
\newacronym{APB}{APB}{Advanced Peripheral Bus}
\newacronym{API}{API}{Application Programming Interface}
\newacronym{ASIC}{ASIC}{Application-Specific Integrated Circuit}
\newacronym{AVX}{AVX}{Advanced Vector Extension}
\newacronym{AXI}{AXI}{Advanced eXtensible Interface}
\newacronym{BLAS}{BLAS}{Basic Linear Algebra Subprograms}
\newacronym{BNN}{BNN}{Binary Neural Network}
\newacronym{CHI}{CHI}{Coherent Hub Interface}
\newacronym{CMOS}{CMOS}{Complementary Metal-Oxide-Semiconductor}
\newacronym{CNN}{CNN}{Convolutional Neural Network}
\newacronym{CPU}{CPU}{Central Processing Unit}
\newacronym{CSR}{CSR}{Control and State Register}
\newacronym{CTS}{CTS}{Clock Tree Synthesis}
\newacronym{DLP}{DLP}{Data Level Parallelism}
\newacronym{DMA}{DMA}{Direct Memory Access}
\newacronym{DRAM}{DRAM}{Dynamic Random-Access Memory}
\newacronym{DSA}{DSA}{Domain-Specific Architectures}
\newacronym{DSP}{DSP}{Digital Signal Processing}
\newacronym{DUT}{DUT}{Device Under Test}
\newacronym{ECL}{ECL}{Emitter-Coupled Logic}
\newacronym{FBB}{FBB}{Forward Body-Biasing}
\newacronym{FDSOI}{FD-SOI}{Fully Depleted Silicon-on-Insulator}
\newacronym{FMA}{FMA}{Fused Multiply-Add}
\newacronym{FPGA}{FPGA}{Field-Programmable Gate Array}
\newacronym{FP}{FP}{Floating Point}
\newacronym{FPU}{FPU}{Floating Point Unit}
\newacronym{GEMM}{GEMM}{General Matrix Multiplication}
\newacronym{GPGPU}{GPGPU}{General-Purpose \acrlong{GPU}}
\newacronym{GPU}{GPU}{Graphics Processing Unit}
\newacronym{HDL}{HDL}{Hardware Description Language}
\newacronym{HERO}{HERO}{Heterogeneous Embedded Research Platform}
\newacronym{HPC}{HPC}{High-Performance Computing}
\newacronym{ILP}{ILP}{Instruction Level Parallelism}
\newacronym{IoT}{IoT}{Internet-of-Things}
\newacronym{IOT}{IoT}{Internet-of-Things}
\newacronym{IPC}{IPC}{Instructions Per Cycle}
\newacronym{IPU}{IPU}{Image Processing Unit}
\newacronym{ISA}{ISA}{Instruction Set Architecture}
\newacronym{LSB}{LSB}{Least Significant Bit}
\newacronym{LSU}{LSU}{Load/Store Unit}
\newacronym{LUT}{LUT}{Look-Up Table}
\newacronym{LVT}{LVT}{low voltage threshold}
\newacronym{MAC}{MAC}{Multiply and Accumulate}
\newacronym{MatMul}{MatMul}{Matrix Multiplication}
\newacronym{MIMD}{MIMD}{multiple instruction, multiple data}
\newacronym{MMU}{MMU}{Memory Management Unit}
\newacronym{MUL}{MUL}{multiplier}
\newacronym{ML}{ML}{Machine Learning}
\newacronym{MVL}{MVL}{maximum vector length}
\newacronym{NUMA}{NUMA}{non-uniform memory access}
\newacronym{NOC}{NoC}{Network-on-Chip}
\newacronym{PCIe}{PCIe}{Peripheral Component Interconnect Express}
\newacronym{PC}{PC}{Program Counter}
\newacronym{PE}{PE}{processing element}
\newacronym{PL}{PL}{Programmable Logic}
\newacronym{PMCA}{PMCA}{Programmable Manycore Accelerator}
\newacronym{PPA}{PPA}{Power, Performance, Area}
\newacronym{PQ}{PQ}{Product Quantization}
\newacronym{PSL}{PSL}{Power Service Layer}
\newacronym{PTE}{PTE}{page-table entry}
\newacronym{PTW}{PTW}{page-table walker}
\newacronym{PULP}{PULP}{Parallel Ultra Low Power}
\newacronym{RAW}{RAW}{read-after-write}
\newacronym{RBB}{RBB}{Reverse Body-Biasing}
\newacronym{ROB}{ROB}{Reorder Buffer}
\newacronym{RTL}{RTL}{Register Transfer Level}
\newacronym{RVT}{RVT}{Regular Voltage Threshold}
\newacronym{RoCC}{RoCC}{Rocket Custom Coprocessor Interface}
\newacronym{SCM}{SCM}{Storage Class Memory}
\newacronym{SEW}{SEW}{Single Element Width}
\newacronym{SIMD}{SIMD}{single instruction, multiple data}
\newacronym{SIMT}{SIMT}{single instruction, multiple thread}
\newacronym{SLDU}{SLDU}{Slide Unit}
\newacronym{SLVT}{SLVT}{super-low voltage threshold}
\newacronym{SM}{SM}{Streaming Multiprocessor}
\newacronym[longplural={Static Random-Access Memories}]{SRAM}{SRAM}{Static Random-Access Memory}
\newacronym{SSE}{SSE}{Streaming SIMD Extension}
\newacronym{SVE}{SVE}{Scalable Vector Extension}
\newacronym{TLP}{TLP}{Thread Level Parallelism}
\newacronym{TxnID}{TxnID}{Transaction ID}
\newacronym{VAC}{VAC}{Vector Access}
\newacronym{VC}{VC}{virtual channel}
\newacronym{VCONV}{VCONV}{Vector Conversion}
\newacronym{VEX}{VEX}{Vector Execute}
\newacronym{VFU}{VFU}{vector functional unit}
\newacronym{VID}{VID}{Vector Instruction Decode}
\newacronym{VIS}{VISSUE}{Vector Instruction Issue}
\newacronym{VLIW}{VLIW}{Very Long Instruction Word}
\newacronym{VLOOP}{VLOOP}{Vector Loop}
\newacronym{VLR}{VLR}{vector length register}
\newacronym{VLSU}{VLSU}{Vector Load/Store Unit}
\newacronym{VNB}{VNB}{Von Neumann Bottleneck}
\newacronym{VRF}{VRF}{Vector Register File}
\newacronym{VT}{VT}{vector thread}
\newacronym{BW}{BW}{bandwidth}
\newacronym{MASKU}{MASKU}{Mask Unit}
\newacronym{VU0.5}{VU0.5}{Vector Unit 0.5}
\newacronym{VU1.0}{VU1.0}{Vector Unit 1.0}
\newacronym{VMFPU}{VMFPU}{Vector Multiplier/Floating Point Unit}
\newacronym{VFPU}{VFPU}{Vector Floating Point Unit}
\newacronym{VDIV}{VDIV}{Vector Divider}
\newacronym{VMUL}{VMUL}{Vector Multiplier}
\newacronym{WAR}{WAR}{write-after-read}
\newacronym{WAW}{WAW}{write-after-write}
\newacronym{DCT}{DCT}{discrete cosine transform}
\newacronym{TSV}{TSV}{through-silicon via}
\newacronym{3DIC}{3D-IC}{three-dimensional integrated circuit}
\newacronym{F2F}{F2F}{face-to-face}
\newacronym{IC}{IC}{integrated circuit}
\newacronym{C4}{C4}{controlled collapse chip connection}
\newacronym{FEOL}{FEOL}{front end of the line}
\newacronym{BEOL}{BEOL}{back end of the line}
\newacronym{SLEN}{SLEN}{striping distance}
\newacronym{VSU}{VSU}{Vector Store Unit}
\newacronym{DNN}{DNN}{Deep Neural Networks}
\newacronym{AI}{AI}{Artificial Intelligence}
\newacronym{AR}{AR}{Augmented Reality}
\newacronym{SoA}{SoA}{State-of-the-Art}
\newacronym{FD-SOI}{FD-SOI}{Fully Depleted - Silicon on Insulator}
\newacronym{RVV}{RVV}{RISC-V ``V''}
\newacronym{ALU}{ALU}{Arithmetic-Logic Unit}
\newacronym{FFT}{FFT}{Fast Fourier Transform}
\newacronym{VLDU}{VLDU}{Vector Load Unit}
\newacronym{ANN}{ANN}{approximate nearest neighbour search}
\newacronym{STE}{STE}{straight-through estimator}

\begin{document}
\title{Stella Nera: A Differentiable Maddness-Based Hardware Accelerator for Efficient Approximate Matrix Multiplication}
\makeatletter
\newcommand{\linebreakand}{%
  \end{@IEEEauthorhalign}
  \hfill\mbox{}\par
  \mbox{}\hfill\begin{@IEEEauthorhalign}
}
\makeatother
\ifdefined\blind
\author{\IEEEauthorblockN{Anonymous}
\IEEEauthorblockA{\textit{Anonymous} \\
\textit{Anonymous}\\
Anonymous}
\and
\IEEEauthorblockN{Anonymous}
\IEEEauthorblockA{\textit{Anonymous} \\
\textit{Anonymous}\\
Anonymous}
\and
\IEEEauthorblockN{Anonymous}
\IEEEauthorblockA{\textit{Anonymous} \\
\textit{Anonymous}\\
Anonymous}
\linebreakand
\IEEEauthorblockN{Anonymous}
\IEEEauthorblockA{\textit{Anonymous} \\
\textit{Anonymous}\\
Anonymous}
\and
\IEEEauthorblockN{Anonymous}
\IEEEauthorblockA{\textit{Anonymous} \\
\textit{Anonymous}\\
Anonymous}
}
\else
\author{\IEEEauthorblockN{Jannis Sch\"onleber}
\IEEEauthorblockA{\textit{Integrated Systems Laboratory} \\
\textit{ETH Zurich}\\
janniss@iis.ee.ethz.ch}
\and
\IEEEauthorblockN{Lukas Cavigelli}
\IEEEauthorblockA{Computing Systems Lab \\
\textit{Zurich Research Center, Huawei Technologies}\\
lukas.cavigelli@huawei.com}
\and

\IEEEauthorblockN{Matteo Perotti}
\IEEEauthorblockA{\textit{Integrated Systems Laboratory} \\
\textit{ETH Zurich}\\
mperotti@iis.ee.ethz.ch}
\linebreakand
\IEEEauthorblockN{Luca Benini}
\IEEEauthorblockA{\textit{Integrated Systems Laboratory} \\
\textit{ETH Zurich and University of Bologna}\\
lbenini@iis.ee.ethz.ch}
\and
\IEEEauthorblockN{Renzo Andri}
\IEEEauthorblockA{Computing Systems Lab \\
\textit{Zurich Research Center, Huawei Technologies}\\
renzo.andri@huawei.com}

}
\fi

\maketitle

\thispagestyle{fancy}
\fancyhf{}
\renewcommand{\headrulewidth}{0pt}
\renewcommand{\footrulewidth}{0pt}

\fancyfoot[C]{\textcopyright 2025 IEEE.  Personal use of this material is permitted.  Permission from IEEE must be obtained for all other uses, in any current or future media, including reprinting/republishing this material for advertising or promotional purposes, creating new collective works, for resale or redistribution to servers or lists, or reuse of any copyrighted component of this work in other works.}

\begin{abstract}

Artificial intelligence has surged in recent years, with advancements in machine learning rapidly impacting nearly every area of life. However, the growing complexity of these models has far outpaced advancements in available hardware accelerators, leading to significant computational and energy demands, primarily due to matrix multiplications, which dominate the compute workload. \textsc{Maddness} (i.e., Multiply-ADDitioN-lESS) presents a hash-based version of product quantization, which renders matrix multiplications into lookups and additions, eliminating the need for multipliers entirely. We present \textsc{Stella Nera}\footnotemark, the first \textsc{Maddness}-based accelerator achieving an energy efficiency of 161\,TOp/s/W@0.55V, 25x better than conventional MatMul accelerators due to its small components and reduced computational complexity. We further enhance \textsc{Maddness} with a differentiable approximation, allowing for gradient-based fine-tuning and achieving an end-to-end performance of 92.5\% Top-1 accuracy on CIFAR-10.
\end{abstract}


\section{Introduction}
\footnotetext{Named after \ifdefined\blind \textit{hidden for double-blind review} \else an album of the famous Swiss musicians Patent Ochsner\fi}

\isvlsi{Machine learning (ML) has advanced rapidly, achieving human-level performance. However, hardware energy efficiency improves only $\sim$1.45$\times$ per technology generation ($\approx$\,3 years) \cite{tsmc_infos}, far slower than the yearly growth in model complexity. This gap makes it increasingly difficult to run state-of-the-art (SoA) AI models on energy-constrained devices like mobiles and IoT nodes.}

\isvlsi{To address this challenge, academia and industry have pursued hardware-algorithm co-design to efficiently run ML workloads. Most accelerators focus on convolution, matrix multiplication, and multiply-accumulate (MAC) operations as core building blocks. Due to their generality and small kernel sizes, accelerating matrix multiplications has become a standard approach in industrial products (e.g., Nvidia Volta, Huawei Ascend \cite{liao2019davinci}). Specialized engines (e.g., im2col) transform data layouts to compute convolutions via matrix multiplications, offering a strong area-performance trade-off. Many accelerators also leverage quantization, pruning, and sparsity exploitation \cite{dac2023, dac2023_2, dac2023_3, dac2022} to reduce datapath width, logic, and memory requirements, achieving significant energy efficiency gains \cite{Reuther_2022, 9415634}. However, like Moore's Law, quantization is reaching its limits too \cite{8351807,9401214}, e.g., \glspl{BNN} encode their weights into bits.}

\isvlsi{Multiplication of low-bit formats can be implemented as simple lookups, but also leads to many lookups. An alternative solution is vectorized multiplication lookup, where the multiplication becomes a linear interpolation of vector lookups of prototypes. The number of prototypes can be significantly reduces, when one of the matrices is constant, which is the case for the weights in the inference case.}

Furthermore, focusing on non-linear quantization enables a better representation of values. For example, \cref{fig:valuevsdensity} shows the histogram of input values (i.e., at Layer 2 of ResNet-9 using Maddness algorithm), by using uniform quantization (indicated by the green crosses) and 4 bits, where only a few quantized values represent a large percentage of the actual values, but many quantized values encode significantly less frequent numbers. Instead, the red dotted lines show the learned patterns by Maddness representing much closer the actual value distribution.

\textsc{Maddness} \cite{blalock2021multiplying} leverages this idea, and uses a hashed version of \gls{PQ} to find closest vector representation. The input representation is determined through a balanced binary regression tree, which leads to \gls{LUT} index. By reading the LUT, we directly get the partial sum of the dot product of the corresponding encoding.
This process replaces all costly \gls{MAC} operations with inexpensive additions, simplifying the overall \gls{MatMul} to shifts, adds, and comparisons, paving the way for efficient hardware acceleration. 

This work makes the following contributions:
\begin{enumerate}
    \item We present \textsc{Stella Nera}, an energy efficiency-optimized hardware architecture implementing the \textsc{Maddness} algorithm. 
    \item An open-source and fully parameterized hardware implementation with the corresponding evaluations and post-layout simulations, demonstrating an energy efficiency of up to 43.1\,TOp/s/W@0.55V with a throughput of 2.9\,TOp/s@0.55V in a competitive commercial 14nm technology.
    \item We introduce the first formulation of \textsc{Maddness} that is differentiable, allowing the method to be used for \gls{DNN} training with drop-in \textsc{Maddness} replacements for Linear and Conv2D layers for PyTorch.
    \item We show end-to-end results for ResNet-9 \cite{resnet} on CIFAR-10 reaching an accuracy of 92.6\%, which is just an accuracy drop of 1.2\% compared to its full-precision baseline, a major step forward over previous results. 
\end{enumerate}

All the results, RTL implementation, verification tests, algorithm implementation, PyTorch layers, training scripts, and artifacts are available open-source at \githublink.
\begin{figure}
    \centering
    \includegraphics[trim={0cm 2cm 8cm 0.4cm},width=0.9\linewidth]{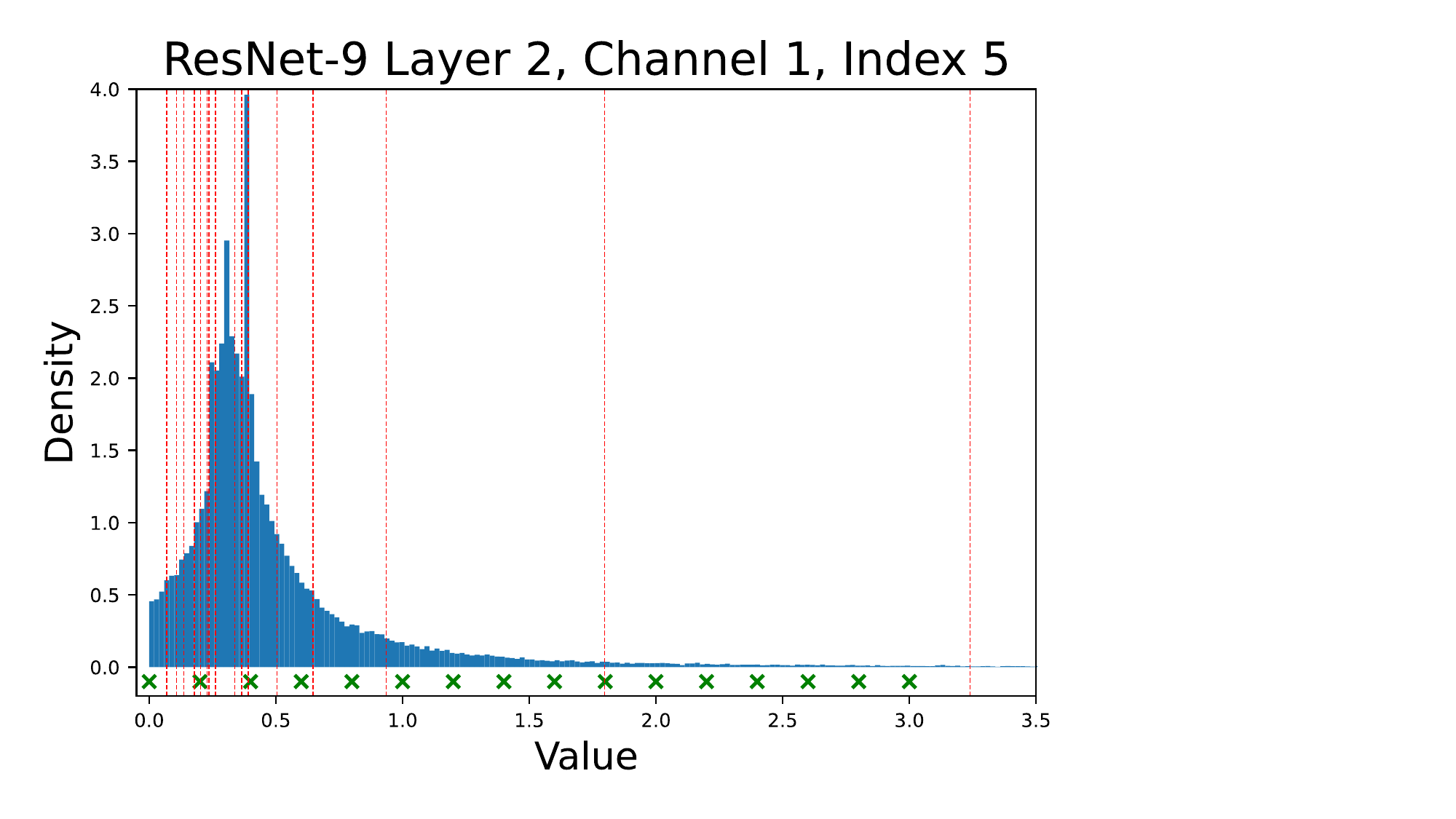}
    \caption{Sample histogram of input feature maps and non-linear quantization levels (red lines) vs. linear levels (green crosses).}
    \label{fig:valuevsdensity}
\end{figure}

\section{Related Work}
Even though the need for efficient circuits for multiplications has never been as important as it is today, trading precision for better energy and area efficiency has been researched for decades. 
Truncation of some least or low significant bits is a popular approach. \cite{kulkarni2011trading} presented multiplier, which also features a precision-scalable and error-aware correction method to compensate for the precision loss. Instead, logarithm-based multipliers replace multiplication by addition but require the application of a logarithm to the operands.

\gls{AMM} typically reduces computation by projecting matrices into lower-dimensional spaces before performing an exact multiplication. A common approach to finding good projection matrices is using one of the many sketching algorithms: either deterministic variants based on \cite{liberty2013simple}, randomized variants, or using sampling techniques \cite{maddness}. These sketching methods consider each matrix in isolation and focus on large matrices.

\isvlsi{Hashing is a common approach to transforming data of any size into a fixed-size value, deploying a hash function (e.g., for efficient data retrieval). Hash function can also be used to calculate vector products (i.e., product quantization).}

\isvlsi{Significant advances have been proposed on the encoding side and a comprehesinve overview can be found in \cite{pqsurvey}.}
\isvlsi{D. Blalock et al. proposed a fast and efficient encoding function based on the Bolt method \cite{blalock2017bolt} by reducing the number of centroids and increasing the number of subvectors compared to previous PQ implementations. However, the similarity function is still norm-based. \textsc{Maddness} futher replaces the $\ell_2$-norm similarity function ($\operatorname{\argmin}$) during the encoding process by a balanced binary regression tree. }

\isvlsi{\textbf{PQ-based methods for DNNs} T. Chen et al. \cite{chen} was the first to propose backpropagation by utilizing the hard $\operatorname{\argmax}$ in the forward path and the differentiable $\mathrm{softmax}$ in the backward path to compress embeddings. PECAN \cite{ran2022pecan} introduce the $\ell_1$-norm-based similarity measure to an entire CNN. In LUT-NN \cite{tang2023lutnn}, and Carter et al. \cite{amazoncarter} further improved the training methods for end-to-end replaced networks and extended it to BERT, in addition to CNNs, while still using a norm-based encoding function limiting the potential speed-up as one moves Ops from the \gls{MatMul} to prototype matching in contrast to \textsc{Maddness} where the ceiling is significantly higher.}

\isvlsi{Chen et al. and Spring et al. \cite{chen2020slide, chen2015compressing, spring2016scalable} apply hashing also for dense layers using locality-sensitive hashing to select which neurons are active in the next layer. These methods significantly differ from \textsc{Maddness} as their goal is to compress and, by that, reduce the number of operations compared to approximating the output of the \gls{MatMul}, allowing \textsc{Maddness} to be used more broadly.}

Fernandez-Marques et al. \cite{fernandezmarques2023yet} propose a custom PQ hardware accelerator (PQA) using an HLS-based implementation on FPGA, although not providing any energy numbers. We present the first implementation of a \textsc{Maddness} accelerator.

\subsection{Approximate Matrix Multiplication}
In general, \gls{AMM} aims to implement efficient matrix-multiplications while trading precision for efficiency (i.e., time, area or power). The problem can be described as follows:   
\begin{equation} \label{eq:amm}
\norm{\underbrace{\alpha \cdot dec(enc(\mathbf{A}), l(\mathbf{B})) + \beta}_{\mathbf{A\tilde{\cdot}B}\approx \mathbf{AB}} - \mathbf{AB}}_F < \epsilon(\tau)\norm{\mathbf{A}\mathbf{B}}_F,
\end{equation}
where  $\mathbf{A} \in \mathbb{R}^{N \times D}$ and $\mathbf{B} \in \mathbb{R}^{D \times M}$ are two matrices. $\mathbf{AB}$ is their numerically precise matrix multiplication and $\mathbf{A\tilde{\cdot}B}$ the approximated matrix multiplication. $\norm{\mathbf{A}}_F$ refers to the Frobenius norm 
\begin{equation} \label{eq:frobenius}
\norm{\mathbf{A}}_F = \sqrt{\sum_{i=1}^{N}\sum_{j=1}^{D}\norm{a_{ij}}^{2}}.
\end{equation}

Given $N \gg D \ge M$ and a time budget $\tau$, one has to find the decoding function $dec(\cdot,\cdot)$, the encoding function for $\mathbf{A}$ $enc(\cdot)$, and the encoding function of $\mathbf{B}$ $l(\cdot)$. The original Maddness paper applied Maddness just on the classifier layer of VGG16, which is a fully-connected neural network layer followed by a softmax activation layer. However, convolutions, which contribute the biggest computational share, also fit into this format when using im2col to map the computation to \glspl{MatMul}. Furthermore, \glspl{DNN}' weight matrix $\mathbf{B}$ is known beforehand in the case of inference, a key prerequisite for \textsc{Maddness}. In the following, we will define the normalization parameters as $\alpha = 1$ and $\beta = 0$ as the input data are normalized in batch normalization. To complete the formulation, we assume a training set $\mathbf{\Tilde{A}}$, which is drawn from the same distribution as $\mathbf{A}$.

\subsection{Product Quantization}
 PQ is a vector quantization method to approximate inner products or distances for \gls{ANN}. Conceptually, PQ approximates $\mathbf{a}\mathbf{b} \approx \Tilde{\mathbf{a}}\mathbf{b}$ for small $\norm{\Tilde{\mathbf{a}} - \mathbf{a}}$. $\Tilde{\mathbf{a}}$ consists of prototypes $\mathbf{P}$ that have been learned ahead of time. The prototypes and the known matrix $\mathbf{B}$ are multiplied offline to construct the lookup table $\mathbf{L}$ using $\mathbf{P}$ and $\mathbf{B}$. An encoding function $enc(\mathbf{a}) = \Tilde{\mathbf{a}}$ maps an input vector $\mathbf{a}$ to the corresponding prototypes. The resulting prototype corresponds to an offline-learned $\mathbf{L}$ entry. For each of the $C$ codebooks (i.e. subspaces), the encoding function maps to one of the $K$ prototypes per subspace. During decoding, the encoded entries of $\mathbf{L}$ are summed up to form the result. Vector $\mathbf{a}$ is a row of $\mathbf{A} \in \mathbb{R}^{N \times D}$ and $\mathbf{b}$ is a column of $\mathbf{B} \in \mathbb{R}^{D \times M}$. The essential components of PQ:
\begin{itemize}
    \item \textbf{Prototypes P} The Prototypes are learned offline. The input matrix $\Tilde{\mathbf{A}}$ is split into $C$ disjoint subspaces of size $\frac{D}{C}$ and for each subspace $K$ prototypes are learned using K-means algorithm. 
    From the dataset, the corresponding $\mathbf{A}$ matrices are sampled and in the following indicated by $\Tilde{\mathbf{A}}$. Define $\mathbb{I}_c = \{0,1,...,\frac{D}{C}-1\}$ as the indices used in every subspace $c$. The subspaces created by these indices need to be disjoint. Learning prototypes $\mathbf{z}_{c,k} \in \mathbb{R}^{|\mathbb{I}_c|}$ or prototype space $\mathbf{P} \in \mathbb{R}^{C\times K \times |\mathbb{I}_c|}$ means minimizing
    \begin{equation}
        \label{eq:offline_learning}
        \sum_{n=1}^{N}\sum_{c=1}^{C}\sum_{i \in \mathbb{I}_c} (\Tilde{\mathbf{A}}_{ni} - \mathbf{z}_{c,k,i} )^{2}. 
    \end{equation}
    One can achieve that by running K-means in every subspace $\mathbb{I}_c$ to learn the cluster mapping $\mathbf{z}_{n, c, i} \in \mathbb{R}^{N} = \mathbf{z}$ and write the resulting centroids into $\mathbf{P}_c \in \mathbb{R}^{|\mathbb{I}_c| \times K}$ where $|\mathbf{\mathbb{I}_c}| = \frac{D}{C}$.\\
\end{itemize}
\begin{figure*}
    \centering
    \includegraphics[width=.85\textwidth]{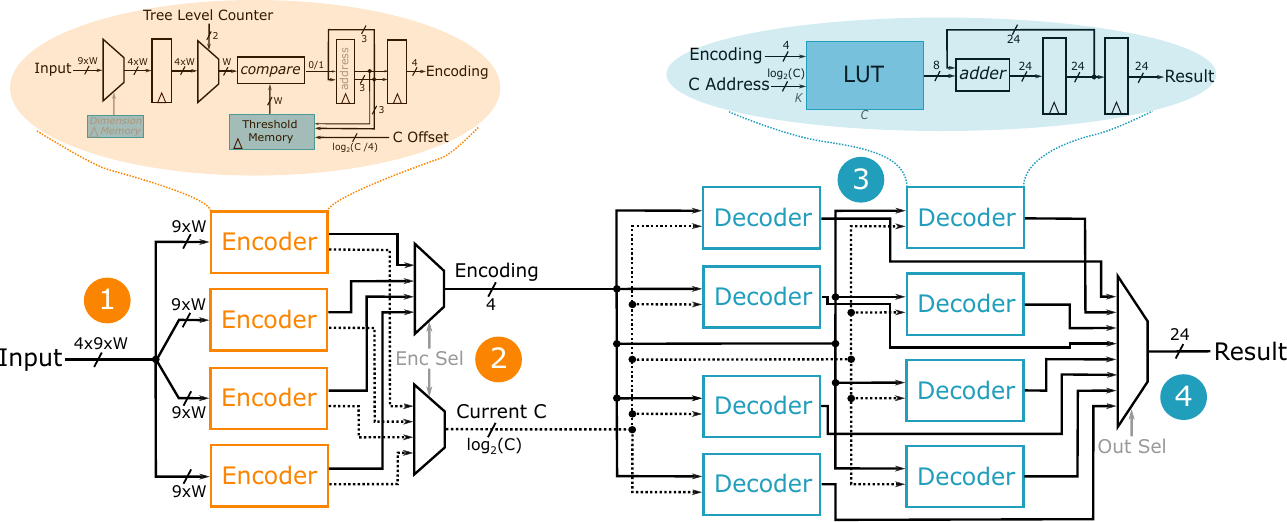}
    \caption{Architecture of \textsc{Stella Nera}: \protect\graphnumbercolor{color5}{1}The input gets split up into the four encoders. The encoders are offset by one cycle and need four cycles to traverse a tree. This leads to one valid output per cycle\protect\graphnumbercolor{color5}{2}. The encoded value and the current $C$ are the \gls{LUT} address\protect\graphnumbercolor{color2}{3}. After C cycles, all decoders have a valid result. A multiplexer selects the results in order\protect\graphnumbercolor{color2}{4}.}
    \label{fig:arch}
    \vspace{-0.5cm}
\end{figure*}
    \begin{itemize}
    \item \textbf{Encoding Function} $enc(\mathbf{A}, \mathbf{P})$\graphnumbercolor{color5}{2}A function that calculates the closest prototype to the row vector $\mathbf{a}$ based on a similarity measurement. This is done for every subspace $C$, resulting in $C$ encoding indices per input row.
    The default similarity measurement $\ell_2$-norm results in the encoding function
    \begin{equation}
        enc_{c}(\mathbf{a}) = \underset{k}{\operatorname{\argmin}} \sum_{i \in \mathbb{I}_c} (\mathbf{a}_i - \mathbf{P}_{c,k,i})^{2}.
    \end{equation}
    \item \textbf{(Subspace) Decoding Function} $\mathbf{L}=lut(\mathbf{B}^{T},\Tilde{\mathbf{P}})  \in \mathbb{R}^{C \times K \times M}$\graphnumbercolor{color2}{4} is computed offline using the determined prototypes $\mathbf{P} \in \mathbb{R}^{C\times K \times \frac{D}{C}}$ and weight matrix $\mathbf{B} \in \mathbb{R}^{D \times M}$.
    Let $\Tilde{\mathbf{P}} \in \mathbb{R}^{C \times K \times D}$ be the expanded version of $\mathbf{P}$ where one adds 0 for all indices in 1\textsuperscript{st} dimension that aren't in the corresponding subspace $i \not\in \mathbb{I}_c$. 

    \begin{equation}
        \mathbf{L}_{c,k,m} = \sum_{i=1}^{D}\mathbf{B}_{m, i}\mathbf{P}_{c, k, i}.
        \label{eq:lut}
    \end{equation}
    \item \textbf{Accumulation} $dec(enc(\mathbf{a}, \mathbf{P}), \mathbf{L})$\graphnumbercolor{color2}{5} implements the superposition of the $C$ subspaces, implemented as accumulation of the corresponding $C$ lookup table entries with the determined indices from the encoding step,
    \begin{equation}
        (\mathbf{AB})_{n,m} \approx \sum_{c=1}^{C}\mathbf{L}_{m, c, k}, \  k = enc_{c}(\mathbf{a}_n).
        \label{eq:accumulation}
    \end{equation}
\end{itemize}

\subsection{\textsc{Maddness}}
\label{sec:maddness}
Product quantization profits mostly when $N,M\gg D$. In the case of one of the matrices fixed, (e.g., with constant weights in the inference), this can be relaxed to $N\gg M,D$, but increases the complexity of the encoding function $enc_{c}(\mathbf{a})$. \textsc{Maddness} find the most similar prototype using locality-sensitive hashing instead of calculating the Euclidian distance for every sub-vector and each prototype. Thus, every sub-vector is hashed to one of the K buckets, and similar vectors end up in the same bucket.

\isvlsi{To achieve efficient training with minimal computation, existing hash functions were not suitable. Instead, \textsc{Maddness} deploys trainable hash functions using balanced binary regression trees, where each leaf node represents a unique hash bucket. The tree selects a leaf for a vector $x$ by traversing from the root: moving left if a specific vector component $x_j$ is below a node-specific threshold $v$, and moving right otherwise. }

\begin{figure}
    \centering
    \includegraphics[trim={1cm 1.5cm 10cm 0},width=0.90\linewidth]{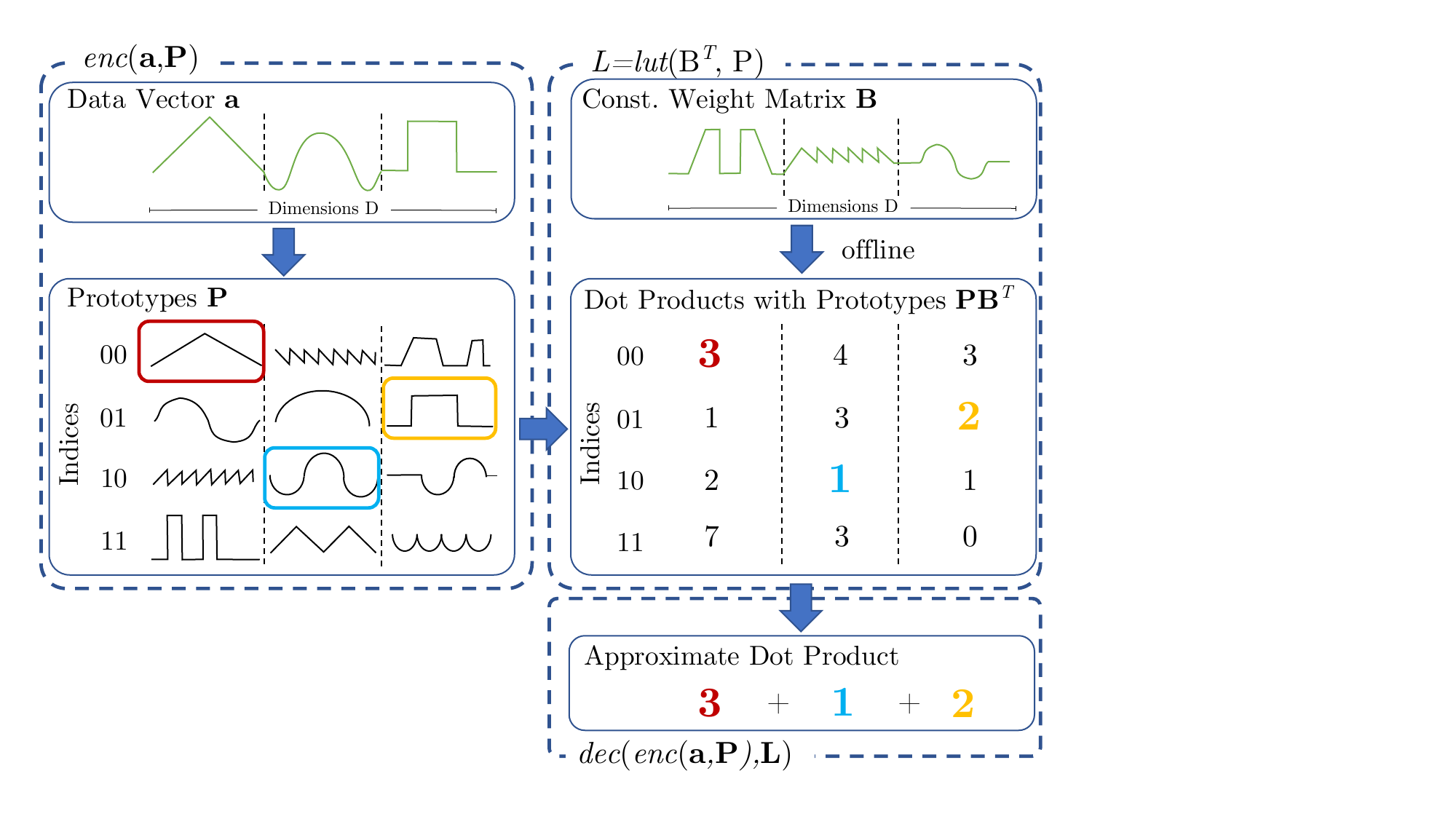}
    \caption{\textsc{Maddness} multiplication. 1) Calibration/Training: prototypes $\mathbf{P}$ are learned to approximate input row vectors $\mathbf{a}$. The encoding function maps $\mathbf{a}$ to the closest prototype. Dot products of the propotoypes with the weights $\mathbf{B}$ are precomputed and stored in a LUT. 2) Inference: Partial dot products are retrieved from LUT and accumulated for approximate matrix mult. \cite{maddness} }
    \label{fig:maddnessdetail}
    \vspace{-0.5cm}
\end{figure}

The algorithm is illustrated conceptually in 
\cref{fig:maddnessdetail} and more in detail in \cref{fig:arch}.
\textsc{Maddness} replaces two components of PQ: the encoding function $enc_{c}(\mathbf{a})$ and the values stored in the LUT $\mathbf{L}$ (marked as \graphnumbercolor{color2}{4} in \cref{fig:arch}). A balanced binary tree (marked as \graphnumbercolor{color5}{2}) replaces the hash function and the vector product between the prototypes $\mathbf{P}$ and the matrix $\mathbf{B}$ are stored inside $\mathbf{L}$. Using a hash function instead of the $l_2$-norm (i.e., $\norm{a-b}_2$) results in a significant speedup with a slight accuracy loss. This allows \textsc{Maddness} to be used on large input matrices $\mathbf{A}$ as every input row has to be encoded.

\subsection{Differentiable \textsc{Maddness}}
A major limitation of the hash function is its non-differentiable nature, rendering it unsuitable for gradient-based optimizations. Based on the Zhang et al.'s on decision trees \cite{zhang2022representation}:

Let \(\mathbf{S}_{c} \in \{0, 1\}^{(K-1) \times \log_2(K)}\) be the selection matrix that describes the structure of the tree. Let \(\mathbf{H}_{c} \in \{-1, 0, 1\}^{K \times (K-1)}\) be the ternary tree description matrix that indicates the direction of decisions, where \(K\) represents the number of prototypes. For simplicity, the following formulas and figures illustrate the two-level decision tree:
\begin{figure}[h]
    \centering
    \begin{tikzpicture}[level/.style={sibling distance=16mm}, level 3/.style={sibling distance=8mm},level 2/.style={sibling distance=8mm},level 4/.style={sibling distance=8mm},
  mtxw/.style = {
    matrix of math nodes,
    nodes in empty cells,
    left delimiter = {[},
    right delimiter = {]},
    every node/.style = {
      anchor = base,
      text width = 1.5em,
      text height = 1.2ex,
      align = right,
      anchor = base east
    }
  },
   mtxuw/.style = {
    matrix of math nodes,
    nodes in empty cells,
    left delimiter = {[},
    right delimiter = {]},
    every node/.style = {
      anchor = base,
      text width = 2em,
      text height = 1.2ex,
      align = right,
      anchor = base east
    }
  },
  mtx/.style = {
    matrix of math nodes,
    nodes in empty cells,
    left delimiter = {[},
    right delimiter = {]},
    every node/.style = {
      anchor = base,
      text width = 1.3em,
      align = center,
      anchor = base east
    }
  },
  mtxs/.style = {
    matrix of math nodes,
    nodes in empty cells,
    left delimiter = {[},
    right delimiter = {]},
    every node/.style = {
      anchor = base,
      text width = 0.5em,
      align = center,
      anchor = base east
    }
  },
  hltr/.style = {opacity = 0.2, rounded corners = 2pt, inner sep = -1pt},
  hltrB/.style = {fill opacity = 0.2, text opacity=1},
  txtup/.style = {rotate = 90, right},
  txtn/.style = {left, xshift = -1em},
  txtnu/.style = {yshift = 3ex},
  txtbt/.style = {yshift = -1ex},
 ]
\useasboundingbox (-4.5,0) rectangle (4,1.0);
\begin{scope}[yshift=0cm, xshift=-2.7cm, scale=0.7]
\matrix (B) [mtxs] {
    $1$ & $0$  \\ 
    $0$ & $1$  \\ 
    $0$ & $1$  \\ 
};
\node[xshift=-0.6cm, scale=0.7] at (B.west) {$\mathbf{S}\enspace= $};
\end{scope}
\begin{scope}[yshift=0cm, xshift=0.5cm, scale=0.7]
\matrix (H) [mtx] {
    $-1$ & $-1$ & $0$ \\
    $-1$ & $1$ & $0$ \\
    $1$ & $0$ & $-1$ \\
    $1$ & $0$ & $1$ \\
};
\node[xshift=-0.6cm, scale=0.7] at (H.west) {$\mathbf{H}\enspace = $};
\end{scope}

\scope[scale=0.5, yshift=1.5cm, xshift=6cm]
\node [draw] (t0){$t_{0}$}
    child {node [draw] (t1) {$t_{1}$}
        child {node (enc0) {0}}
        child {node (enc1) {1}}
    }
    child {node [draw] (t2) {$t_{2}$}
        child {node (enc2) {2}}
        child {node (enc3) {3}}
    };
\endscope

\end{tikzpicture}

\end{figure}

The columns in \(\mathbf{S}_{c}\) represent the number of levels in the decision tree; for our \textsc{Stella Nera} architecture, this value is 4. We are using \(K = 16\) prototypes. The variable \(\mathbf{X}\) in the following formulas refers to the input selected through the \textsc{Maddness} dimension reduction process, specifically \(\mathbf{X} = \{x_1, x_4, x_6, \ldots, x_n\}\) (with four dimensions per codebook). The dimensions were determined by the offline \textsc{Maddness} algorithm (illustrated in \cref{fig:maddnessdetail}).

By subtracting the threshold vector and applying the sign function, the intermediate term of $\sigma(\mathbf{SX}-\theta)$ shows every decision in the decision tree. Multiplying the description matrix with this resulting vector, maps all single decision to the final output class, whereas there is a unique class getting $\log_2(K)$ votes (i.e., guaranteed because the binary tree is balanced), and is set to 1.
\begin{equation}
\mathbf{E}_i = \left\{
\begin{array}{ll}
1 & i= \operatorname{\argmax}_i(\mathbf{H}\sigma(\mathbf{S}\mathbf{X} - \theta))_i \\
-1 & \, \textrm{else} \\
\end{array}
\right. 
    \label{enc:hard}
\end{equation}
where $\sigma$ is the $\mathrm{sign}$ function.

To parallelize over multiple codebooks, let
\begin{align}
    \mathbf{S} &= \bigoplus_{i=1}^C \mathbf{S}_i = \mathrm{diag}(\mathbf{S}_{1}, \mathbf{S}_{2}, .. , \mathbf{S}_{C}), 
\\
    \mathbf{H} &= \bigoplus_{i=1}^C \mathbf{H}_i = \mathrm{diag}(\mathbf{H}_{1}, \mathbf{H}_{2}, .. , \mathbf{H}_{C}).
\end{align}
The decision threshold vector is then $\theta \in \mathbb{R}^{C\times (K-1)}$.

We deploy \eqref{enc:hard} in the forward path. Unfortunately, hard decisions introduce large or infinite gradients, making them unsuitable for SGD-based optimization. We therefore extend it to a soft variant, which we use for back-propagating the gradients only,
\begin{equation}
    \mathbf{E}_\text{soft} = \mathrm{softmax}(\mathbf{H\sigma_{\text{soft}}(\mathbf{SX} - \theta)}),
\end{equation}
using $\sigma_\text{soft} = \mathrm{tanh}$. This method is inspired by the widely used technique called \gls{STE} \cite{bengio2013estimating}.

Using $\mathbf{E} \in \{0, 1\}^{N\times C \times K}$ and the LUT $\mathbf{L}$ (from (\ref{eq:lut})), we first sum over the prototypes $K$, then accumulate over the subspaces $C$: 
\begin{equation}
    \mathbf{I}_{n,m,c} = \sum_{i=1}^{K}\mathbf{E}_{n,c,i}\mathbf{L}_{m,c,i},
    \qquad
    \left(\mathbf{AB}\right)_{n,m} = \sum_{i=1}^{C}\mathbf{I}_{n,m,c}.
\end{equation}

We use this approach to increase accuracy after the offline \textsc{Maddness} initialization or to train a \textsc{Maddness} model starting from a random initialization.

Before applying \textsc{Maddness}, Conv2D layers use the im2col method \cite{im2col} to convert the Conv2D to a \gls{MatMul} using input $\mathbf{X} \in \mathbb{R}^{N \times C_{i} \times H_{i} \times W_{i}}$, weights $\mathbf{W} \in \mathbb{R}^{C_{o} \times C_{i} \times h_{k} \times w_{k}}$, the im2col transformed input $\mathbf{\Tilde{X}} \in \mathbb{R}^{N \cdot H_{i} \cdot W_{i} \times C_{i} \cdot h_{k} \cdot w_{k}}$ is defined as
\begin{equation}
    \begin{aligned}
    \mathbf{\Tilde{X}}(b \cdot N + o_{x} \cdot W_{i} \cdot o_{y}, i_{c} \cdot h_{k} \cdot w_{k} + k_{x} \cdot w_{k} + k_{y}) \\
    = \mathbf{X}(b, i_{c}, o_{x} + k_{x}, o_{y} + k_{x})
\end{aligned}
\end{equation}

for padding and stride equals 1 with $b = 1, \dots, N,\ i_{c} = 1, \dots, C_{i},\ o_{x}\ \mathrm{and}\ o_{y} = 1, \dots, H_{i} \ \mathrm{and}\ W_{i},\ k_{x}\ \mathrm{and}\ k_{y} = 1, \dots, h_{k}\ \mathrm{and}\ w_{k}$. $\mathbf{\Tilde{W}} \in \mathbb{R}^{C_{i} \cdot h_{k} \cdot w_{k} \times C_{o}}$ are the transformed weights. Resulting in the Conv2D being represented as $\mathbf{\mathbf{O} = \mathbf{\Tilde{X}}} \cdot \mathbf{\Tilde{W}}$ a \gls{MatMul} with an inner dimension $D$ of $C_{i} \cdot h_{k} \cdot w_{k}$ allowing us to learn one codebook per input channel when using a codebook width of $h_{k} \cdot w_{k}$. Whereas for kn2col $(H_{i} \cdot W_{i}\times C_{i})\times(C_{i} \times k_{x} \cdot k_{y} \cdot C_{o})$ with an inner dimension of $C_{i}$  using the same codebook-width, keeping the LUT size equal, results in lower accuracy due to the resulting prototypes being across channels. First, we quantize the network to INT8 deploying the Straight-Through Estimator (STE) approach, where quantization is implemented as an identity function in the backward pass, enabling gradient-based optimization. In the forward pass, both activations and weights are quantized to 8 bits.

\section{Architecture}

 \cref{fig:arch} presents the overall architecture of Stella Nera:

\textbf{Encoding Unit} (shown on the top left side in detail) encodes an input vector by iteratively traversing the binary tree $\mathbf{S}$. This takes $log_2(K)$ cycles to generate a valid encoding (i.e., $log_2(K)$ levels of the tree). In our implementation, we use $K=16$, leading to a four cycles latency. Each \textsc{Stella Nera} accelerator implementation has four encoders. These encoders operate with a one-tree-level/cycle offset to produce one valid encoding per cycle together. The depth of the tree (i.e., the number of prototypes $\sqrt{K}$) is a parameter of the architecture. Previous work suggests that $K=16$ generally performs best in the accuracy vs. performance trade-off \cite{tang2023lutnn}.


\textbf{Decoding Unit} (shown on the right): The LUT implemented for the decoder is built with standard cell-based memory (SCM) to allow low-voltage and high-energy efficiency. The decoding unit then processes one encoded value or address into the LUT per cycle. The value obtained from the LUT is then accumulated. After $C$ accumulations (\ref{eq:accumulation}), the output value is the accumulated value. Two variants were implemented, namely, 1) INT8 LUT and accumulation in INT24, and 2) FP16 LUTs and FP32 accumulation. As the LUT is already approximate, the use of INT8 results in a negligible accuracy drop, as also found by Tang et al. \cite{tang2023lutnn}. Therefore, we will only report the results of the INT8/INT24 version.

\textbf{System and Scaling}
When we combine encoder and decoder, as illustrated in \cref{fig:arch}, we get the \textsc{Stella Nera} accelerator. The schematic shows four encoders and eight decoders, with four encoders producing one output per cycle.

\textsc{Stella Nera} has two scaling parameters. The first is the total number of decoders called $N_{dec}$, while the second is the number of codebooks, $C_{dec}$, per decoder. It is important to note that the subparameter $W_{dec}$ encompasses the number of output values per cycle, which is necessary when $N_{dec}$ is greater than $C_{dec}$. In such a case, serving all outputs in the $C_{dec}$ cycles required to calculate the next value would not be possible. The most effective approach is to combine multiple smaller units based on the given problem size. For example, if we have four \textsc{Stella Nera} units with $N_{dec} = 64$ , $C_{dec} = 16$ and $W_{dec} = 8$, we can tile in all three tiling directions: $N, D, M$, providing maximum flexibility. However, an additional adder is required if we tile in the $D$ (i.e., $C$) direction.


\section{Algorithmic Results}


\begin{table*}\centering
\caption{Comparison of state-of-the-art. 1 MAC = 2 Ops.}
\begin{small}

\begin{threeparttable}
\scriptsize
\begin{tabular}{lrrrrrrrr}

\toprule
 & ISSCC \cite{park2022multi} & ISSCC \cite{lin20207} & ISSCC \cite{agrawal20219}  & Nat. Elec. \cite{LeGallo2023} & VLSI Symp. \cite{anders20182} & DATE \cite{redmule}  & \textbf{This work} & \textbf{This work}\tnote{5} \\
\midrule
\textbf{Technology} [nm] & 4 & 7 & 7 & 14 &  14 & 22 & 14 & 3 \\
\hdashline
\textbf{Power Supply} [V] & 0.55 & 0.575 & 0.55 & 0.85 & 0.55 & 0.65 & 0.55 & 0.55\\
\hdashline
\textbf{Precision} & INT8\tnote{1} & INT8\tnote{1} & INT4\tnote{1} & 12b ADC & INT8 & FP16 & INT8 \textit{(LUT)} & INT8 \textit{(LUT)}\\
\hdashline
\textbf{ResNet-9 Acc.} [\%] & 93.6\tnote{7} & 93.6\tnote{7} & 91.7\tnote{7} & 92.8\tnote{6} & 93.6\tnote{7} & \textbf{93.7} & 92.6 & 92.6 \\
\midrule
\textbf{Peak Energy Efficiency} [TOp/s/W]\hspace{-5mm} & 11.6\tnote{2} & 6.8 & 16.5 & 1.7 & 5.2 & 0.7 & \textbf{43.1}\tnote{4} & \textbf{161}\tnote{4} \\
\hdashline
\textbf{Peak Area Efficiency} [TOp/s/mm\textsuperscript{2}]\hspace{-5mm} & 3.45\tnote{2} & \textit{0.4}\tnote{3} & 3.3 & 0.34 & \textit{1.7}\tnote{3} & 0.06 & \textbf{5.1}\tnote{4} & \textbf{163}\tnote{4}  \\
\hdashline
\textbf{Peak Perf.} [TOp/s] & 19.7 & \textit{1.2}\tnote{3} & \textbf{64} & 1.7 & \textit{0.04}\tnote{3} & 0.03 & 2.9\tnote{4} & 4.1\tnote{4} \\
\midrule
\textbf{Area} [mm\textsuperscript{2}] & 4.74 & 3.04 & 19.6 & 11.1\tnote{8} & 0.03 & 0.5 & 0.57 & 0.025 \\ %
\hdashline
\textbf{Frequency} [MHz] & 332 & 290 & 1'000 & - & 400 & 476 & 624 & 886 \\
\hdashline
\textbf{Power} [mW] & 381\tnote{2} & 174 & \textit{3'880}\tnote{3} & - & 8 & 43.5 & 60.9 & 23.0 \\
\bottomrule
\end{tabular}
\begin{tablenotes}
\scriptsize
  \item[1] These chips support additional precisions. \textsuperscript{3} Inferred numbers. \textsuperscript{4} using a codebook width $CW = 9$. \textsuperscript{5} scaled using \cite{deepscaletool} and \cite{tsmc_infos}
  \item[2] End-to-end energy numbers running MobileNetEdgeTPU. Their peak energy efficiency is higher.
  \textsuperscript{6} They use slightly different channel numbers to fit their specific architecture. \textsuperscript{7} Trained via quantization-aware training (QAT) using PyTorch, the code is available online. \textsuperscript{8} Area for 8 cores.
\end{tablenotes}
\end{threeparttable}
\end{small}

\label{tab:soa}
\end{table*}


We evaluated our differentiable implementation of \textsc{Maddness} using the ResNet-9 architecture with the channel numbers $\{64, 128, 128, \allowbreak256, 256, 256, 256\}$, which is a commonly used benchmark architecture for edge scenarios in deep learning. The dataset used for training and evaluation was CIFAR-10. The base model was trained for $N=200$ epochs with a batch size of $128$. 
\begin{figure}
    \centering
    \includegraphics[width=0.9\linewidth]{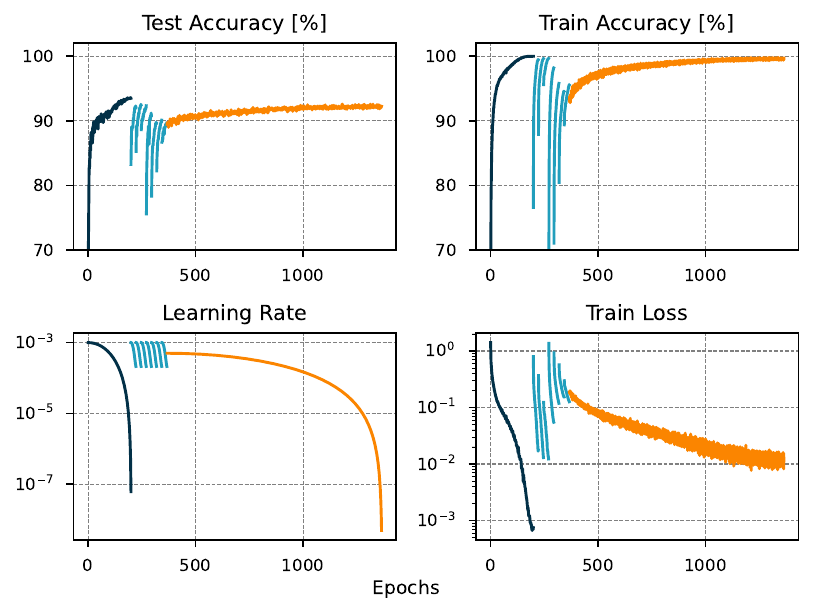}
    \caption{Key metrics for the ResNet-9 training colored with the three different stages \protect\graphnode{color3} pre-training, \protect\graphnode{color2} layer-by-layer and \protect\graphnode{color5} fine-tuning.}
    \label{fig:training}
    \vspace{-2mm}
\end{figure}
For fine-tuning with \textsc{Maddness}, we replace the layers using a codebook width of $CW = 9$ and a number of prototypes $K=16$. This resulted in a LUT table size twice the size of the weights and one unrolled $3\times 3$ kernel per codebook. The replaced layers were initialized using the \textsc{Maddness} algorithm \cite{blalock2021multiplying}. For each layer, the network was trained for $N=25$ epochs with a learning rate $\eta = 10^{-3}$ and half of that learning rate for the thresholds using a cosine annealing scheduler (with $T_{max}=25$, $\eta_{min}=2\cdot 10^{-4}$).


We followed the common practice of calculating the first and last layer in higher-precision (i.e., FP16), because they typically have the highest impact in the total model performance. In our case, these are less than 1\% of compute operations and, therefore, negligible. Once all layers were replaced, the network was trained for $N=1000$ epochs with a learning rate $\eta=5\cdot 10^{-4}$ using a cosine annealing learning rate schedule with $T_{max}=1000$. We employed commonly used data augmentation methods such as random cropping and random horizontal flipping. The pre-training, layer-by-layer conversion with fine-tuning and the final end-to-end fine-tuning stages are visualized in \cref{fig:training}. We obtain an overall accuracy of 92.64\% (see \cref{tab:soa}).

\section{Hardware Implementation and Evaluation}
\textbf{Physical Implementation}
Stella Nera is implemented using industry-standard EDA tools and a 14nm technology. We run SDF back-annotated activity-based post-layout simulation (TT, 0.55V, 25$^{\circ}$C). The energy efficiency is extracted by running a MatMul using randomly generated inputs on a gate-level netlist. The \textsc{Stella Nera} accelerator runs at 624\,MHz (using a maximum of five levels of logic), accounting for hold and setup timing in the respective worst-case conditions. The floorplan of the \textsc{Stella Nera} accelerator, which uses $N_{dec} = 64$, $W_{dec} = 8$, $C_{dec} = 16$, is shown in \cref{fig:floorplan}. As expected, a significant portion of the area is used by LUTs, shown with the\graphnode{color2}numbers. One can also spot the four encoding units in\graphnode{color5} and INT8/INT24 adders in\graphnode{color1}.

\begin{figure}
    \centering
    \includegraphics[width=0.6\linewidth]{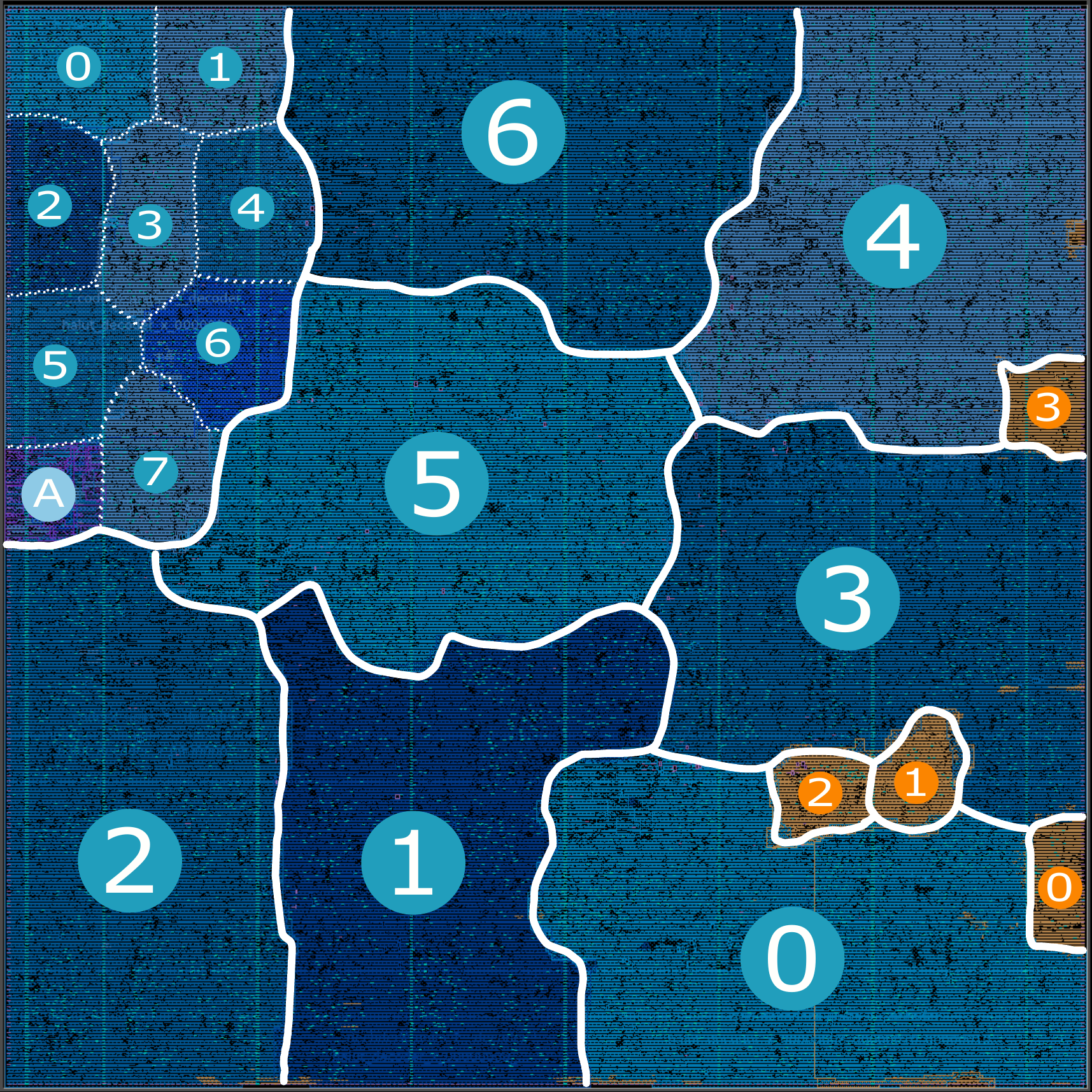}
    \caption{Physical implementation of a \textsc{Stella Nera} accelerator with unit $N_{dec} = 64$, $W_{dec} = 8$, $C_{dec} = 16$ using a commercial 14nm technology. The core area is 350\textmu m$\times$350\textmu m at a utilization of 85\% (860 kGE).}
    \label{fig:floorplan}
    \vspace{-2mm}
\end{figure}

\textbf{Subunit Analysis}
Table \ref{tab:subunit_execute} gives an overview of the power consumption of each subunit of the \textsc{Stella Nera} at full utilization. The energy consumption is dominated by the lookup in the decoding phase (0.26\,pJ per decoder) and the encoding (0.33\,pJ per encoder). The accumulation contributes very little (i.e., 30\,fJ) to the overall power consumption. This means that inexpensive tiling along the $D$ (i.e., $C$) direction of the \gls{MatMul} can be achieved by adding an additional adder when using multiple \textsc{Stella Nera} units. One decoding operation involves $CW$ MACs or $CW \cdot 2$ Ops. The encoding energy is spend once for all decoders $N_{dec}$.

\begin{table}
    \centering
        \caption{Subunit area, power, and energy per operation during the execution of \textsc{Maddness} in 14nm, TT, LVT, 0.55V.}
    \includegraphics[width=1.0\linewidth]{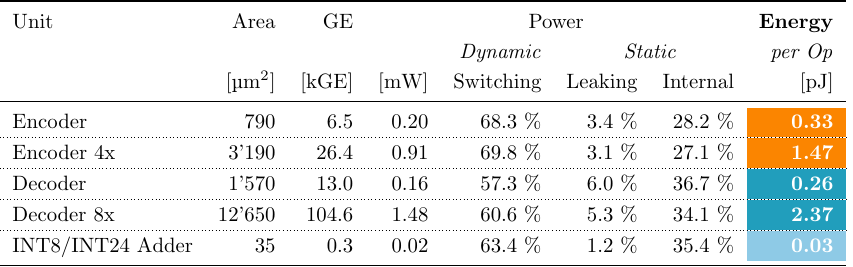}
    \label{tab:subunit_execute}
    \vspace{-5mm}
\end{table}

\textbf{System Results} We combine 4 \textsc{Stella Nera} units, with a total specification of \( N_{dec} = 256 \), \( W_{dec} = 32 \), and \( C_{dec} = 16 \). To create a system capable of running end-to-end an FP16 ResNet-9 or a similar network, we require two FP16 FMA units for each \textsc{Stella Nera} unit. These FP16 FMA units facilitate the conversion of our INT24 results back to FP16. Further, they are utilized for the final small fully-connected layer, which is not accelerated by \textsc{Stella Nera}.

We simulate the system using an FP16 FMA unit \cite{mach2020fpnew}, which can perform two conversions per cycle. When simulating a complete ResNet-9 network with these additional units, we achieve a performance of 984 inferences per second at 23.7 µJ per inference (in 14 nm). Out of this, 13.7\,pJ/cycle or 8.5\,mW (23.3\%) is attributed to the added FMA units. The overall peak energy efficiency is 161 TOP/s/W, which is 13.8 times better than the next best accelerator \cite{park2022multi} at the same voltage and with similar technology.

The bandwidth requirements during calculations are the same as when reading inputs in a conventional Matrix Multiplication, with the weights increased by a factor of \( \frac{16}{9} = 1.77 \times \) due to using \( CW = 9 \) and \( C_{dec} = 16 \). All weights are stored in the LUTs of the \textsc{Stella Nera} units, eliminating the need for traditional weight buffers. The fully configurable system simulation is available online.
In \cref{tab:soa}, we show an overview of state-of-the-art accelerators. We provide the the actual numbers for 14\,nm technology, and further scale the number down to 3nm using DeepScaleTool \cite{deepscaletool} (14nm to 7nm), and foundry-published scale numbers from 7nm to 3nm \cite{tsmc_infos}.

\section{Conclusion}
{\textsc{Stella Nera} is the first hardware implementation of \textsc{Maddness} offering seamless integration into existing AI accelerators for efficient DNN inference. The energy efficiency reaches 43.1\,TOp/s/W@0.55V in 14\,nm or 161\,TOp/s/W@0.55V scaled to 3\,nm with a peak performance of 2.9\,TOp/s and 4.1\,TOp/s, respectively. 15$\times$ higher area efficiency and 25$\times$ higher energy efficiency than the state-of-the-art for application scenarios with comparable accuracy, a significant efficiency improvement.} \isvlsi{Additionally, we enhanced the conversion method for existing DNN models to \textsc{Maddness} by developing a differentiable approximation, enabling gradient-based fine-tuning. This breakthrough eliminates the limitations of converting only few layers in the DNN, allowing us to convert and fine-tune an entire ResNet-9 model. Despite the conversion, \textsc{Maddness} achieves 92.6\% accuracy -- just a 1.1\% drop compared to the full-precision baseline.}
\isvlsi{This work has confirmed the high potential of MADDness-like methods, verified that they can be implemented highly efficiently in hardware. We hope it can serve as an encouragement for further research on similar algorithms and training methods, extending the support to a broader range of models, and overcoming the very long training times. We share all training scripts and RTL implementations on \githublink.}

\bibliography{./bib/paper}
\bibliographystyle{ieeetr}

\end{document}